\documentclass[twocolumn,showpacs,prl]{revtex4}
\usepackage{graphicx}
\usepackage{color}
\usepackage{amssymb}
\newcommand{\lijntje}[3]{(\begin{picture}(20,0)(4,0)
\end{picture})}
\begin{document}
\title{A very broad $X(4260)$ and the resonance parameters\\
of the $\psi (3D)$ vector charmonium state.}
\author{Eef van Beveren}
\email{eef@teor.fis.uc.pt}
\affiliation{Centro de F\'{\i}sica Te\'{o}rica,
Departamento de F\'{\i}sica, Universidade de Coimbra,
P-3004-516 Coimbra, Portugal}
\author{George Rupp}
\email{george@ist.utl.pt}
\affiliation{Centro de F\'{\i}sica das Interac\c{c}\~{o}es Fundamentais,
Instituto Superior T\'{e}cnico, Edif\'{\i}cio Ci\^{e}ncia, Piso 3,
P-1049-001 Lisboa, Portugal}
\author{J.~Segovia}
\email{segonza@usal.es}
\affiliation{Grupo de F\'{\i}sica Nuclear e
Instituto Universitario de F\'{\i}sica Fundamental y Matem\'{a}ticas,
Universidad de Salamanca, 37008 Salamanca, Spain}
\date{\today}

\begin{abstract}
We argue that the $X(4260)$ enhancement contains a wealth of information on
$1^{--}$ $c\bar{c}$ spectroscopy.
We discuss the shape of the $X(4260)$
observed in the OZI-forbidden process $e^{+}e^{-}\to\pi^{+}\pi^{-}J/\psi$,
in particular at and near vector charmonium resonances as well as
open-charm threshold enhancements.
The resulting very broad $X(4260)$ structure
does not seem to classify itself as a $1^{--}$ $c\bar{c}$ resonance,
but its detailed shape allows to identify new vector charmonium states.
Here, we estimate the resonance parameters of the $\psi (3D)$.
\end{abstract}


\maketitle

\begin{figure}[htbp]
\begin{center}
\begin{tabular}{c}
\includegraphics[width=160pt]{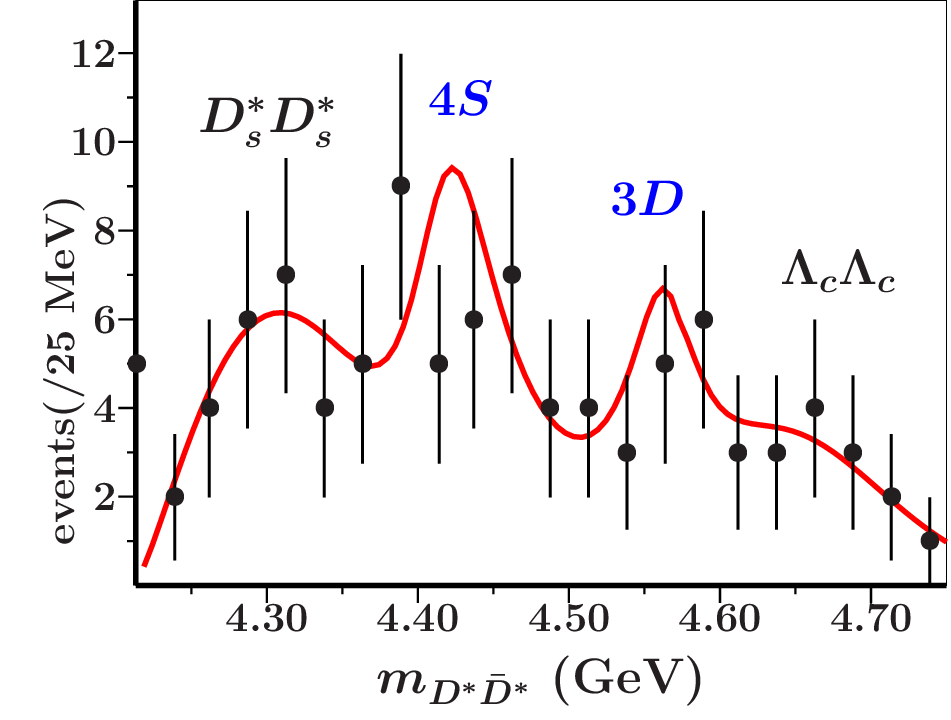}\\ [-20pt]
\end{tabular}
\end{center}
\caption{\small
Event distribution for the reaction
$e^{+}e^{-}\to D^{\ast}\bar{D}^{\ast}$,
as published by the BaBar Collaboration
\cite{PRD79p092001}.
}
\label{B4260b}
\end{figure}

Recent data published by the BaBar Collaboration~\cite{PRD79p092001}
do not exhibit the $X(4260)$ \cite{PLB667p1} structure in
$e^{+}e^{-}\to D^{\ast}\bar{D}^{\ast}$.
However, the data clearly show an enhancement due to the opening
of the $D^{\ast}_{s}\bar{D}^{\ast}_{s}$ channel at 4.213 GeV.
In Fig.~\ref{B4260b} we indicate by a solid line
our interpretation of the data of Ref.~\cite{PRD79p092001}
just above the $D^{\ast}_{s}\bar{D}^{\ast}_{s}$ threshold.
One clearly observes --- albeit with very limited statistics ---
a threshold enhancement, as predicted in Ref.~\cite{AP323p1215},
as well as the two $c\bar{c}$ resonances $\psi(4S)$ and $\psi(3D)$.
The latter charmonium state can be determined from the theoretical model
of Ref.~\cite{PRD21p772}, and was also predicted by Godfrey and Isgur
\cite{PRD32p189}, though a little bit lower, viz.\ at 4.52~GeV.
The $D^{\ast}_{s}\bar{D}^{\ast}_{s}$ threshold enhancement rises fast
and peaks at about 4.32 GeV.
For higher masses,
the threshold signal decreases,
almost vanishing at about 4.75 GeV, where
the $\Lambda^{+}_{c}\Lambda^{-}_{c}$ threshold enhancement
dominates.

The $X(4260)$ $J^{PC}=1^{--}$ charmonium enhancement,
discovered in $\pi^{+}\pi^{-}J/\psi$ by BaBar \cite{PRL95p142001},
was later confirmed and also seen in
$\pi^0\pi^0J/\psi$ as well as $K^{+}K^{-}J/\psi$ by CLEO
\cite{PRL96p162003}, and finally by Belle,
in $\pi^{+}\pi^{-}J/\psi$ \cite{PRL99p182004}, too.
Moreover, both BaBar and
Belle observed a structure in $e^{+}e^{-}\to\pi^{+}\pi^{-}\psi(2S)$ at
somewhat higher energies, namely at 4.32~GeV \cite{PRL98p212001} and
4.36~GeV \cite{PRL99p142002}, respectively.

\refstepcounter{figure}
\begin{figure*}[hbtp]
\begin{center}
\begin{tabular}{ccc}
\includegraphics[height=100pt]{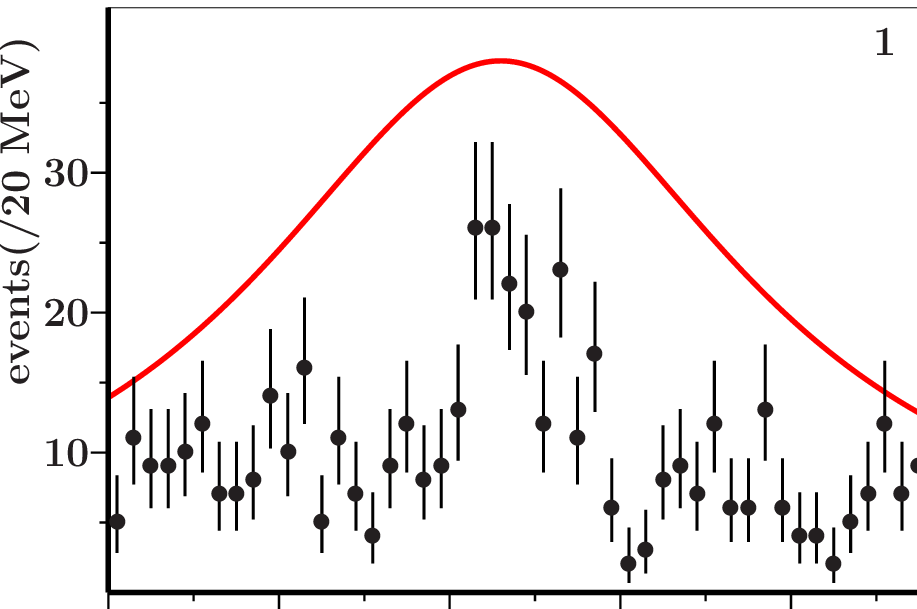} &
\includegraphics[height=100pt]{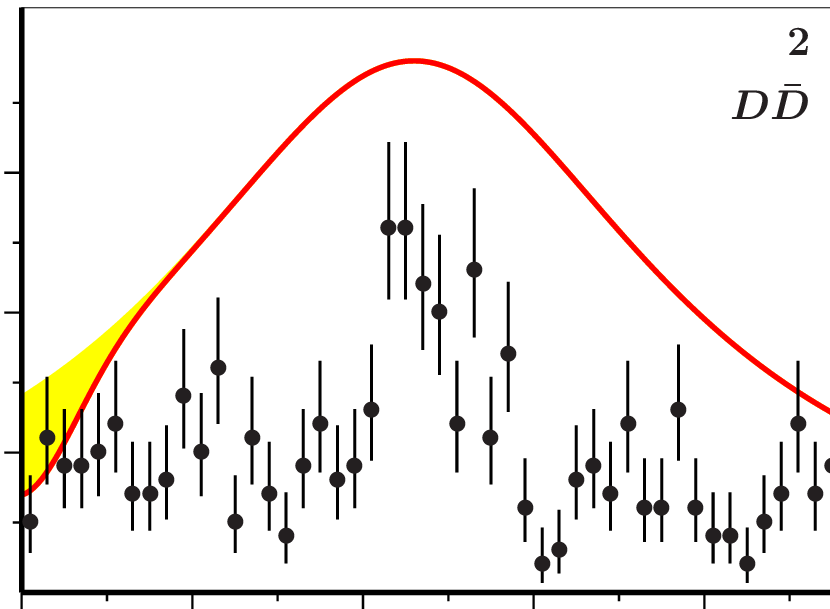} &
\includegraphics[height=100pt]{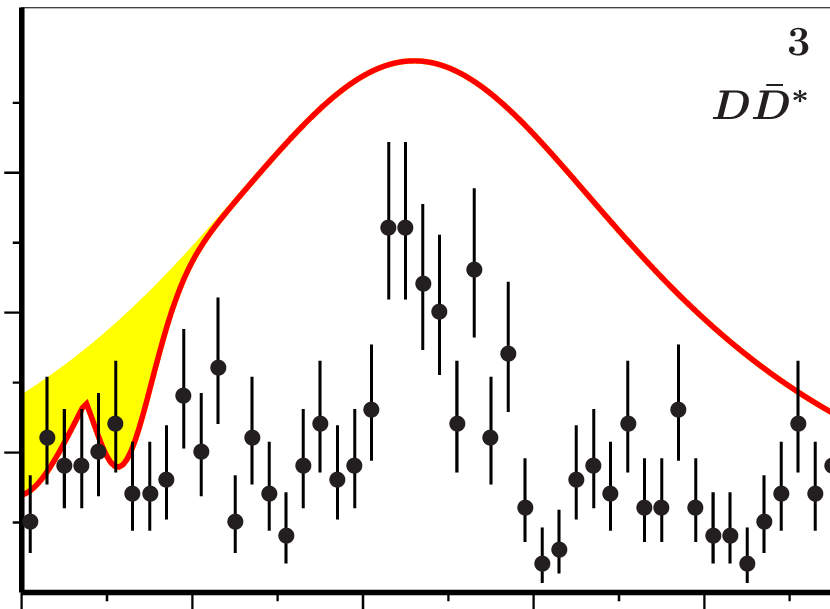} \\
\includegraphics[height=100pt]{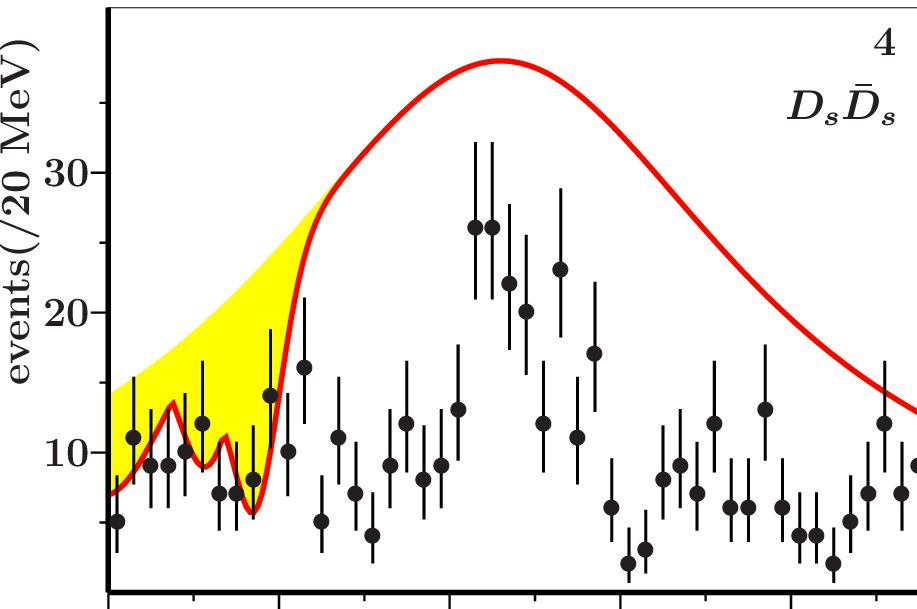} &
\includegraphics[height=100pt]{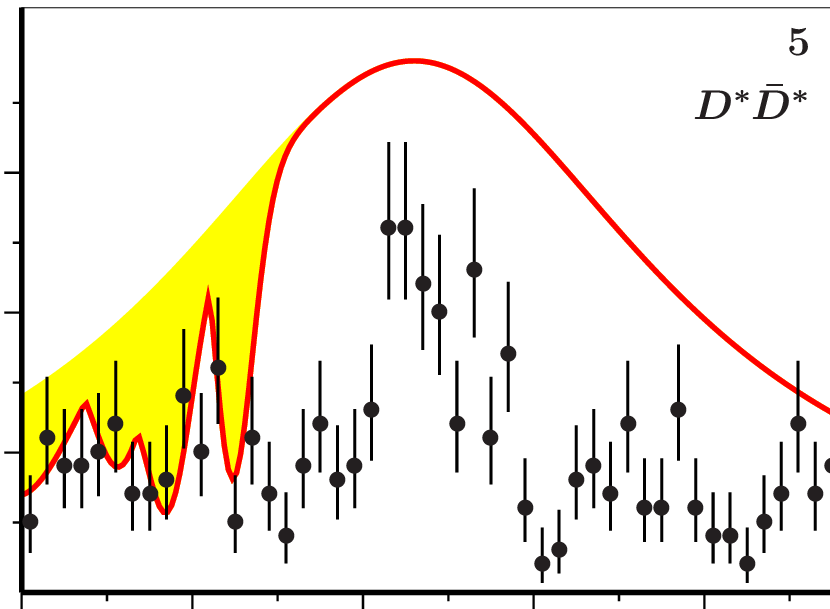} &
\includegraphics[height=100pt]{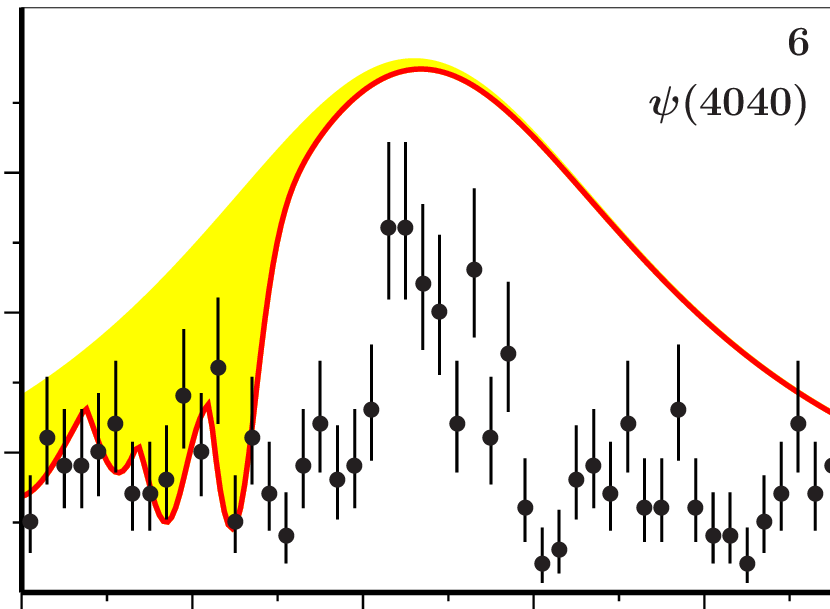} \\
\includegraphics[height=100pt]{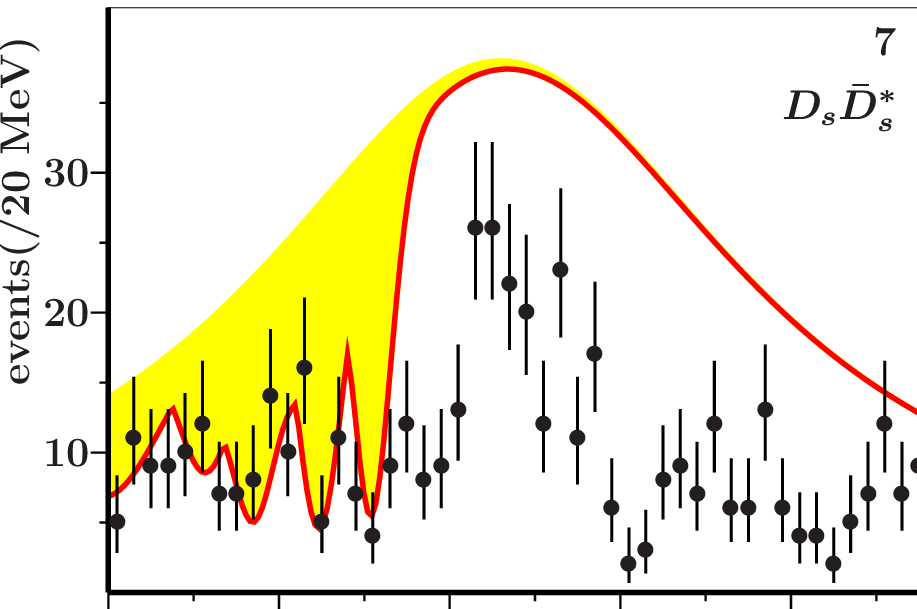} &
\includegraphics[height=100pt]{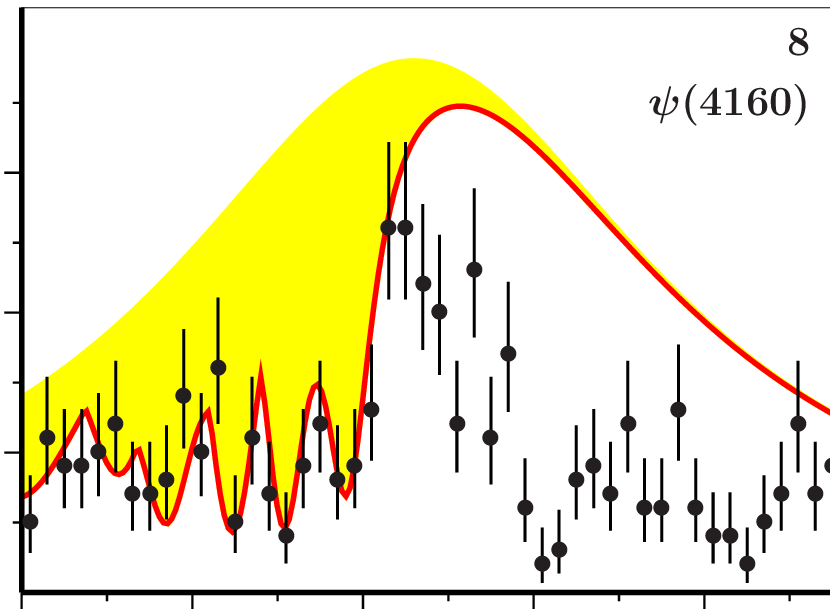} &
\includegraphics[height=100pt]{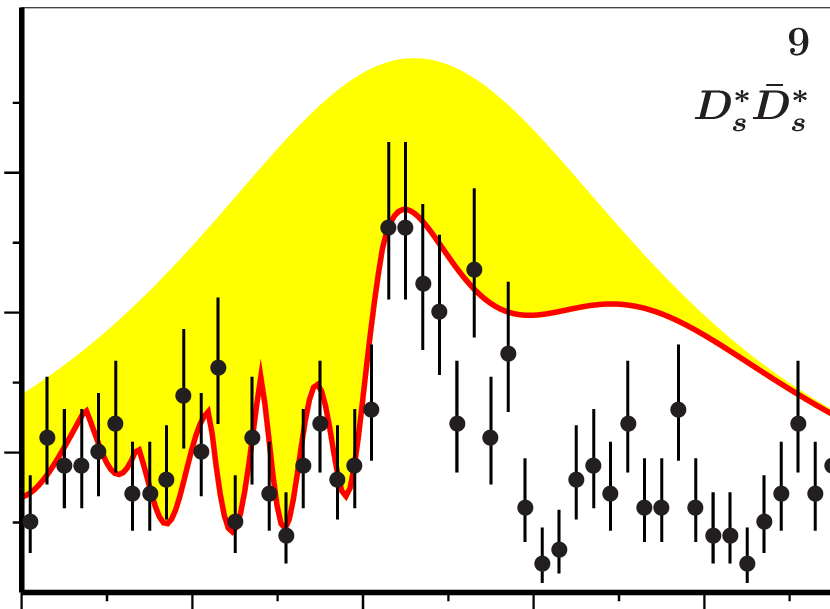} \\
\includegraphics[height=116.48pt]{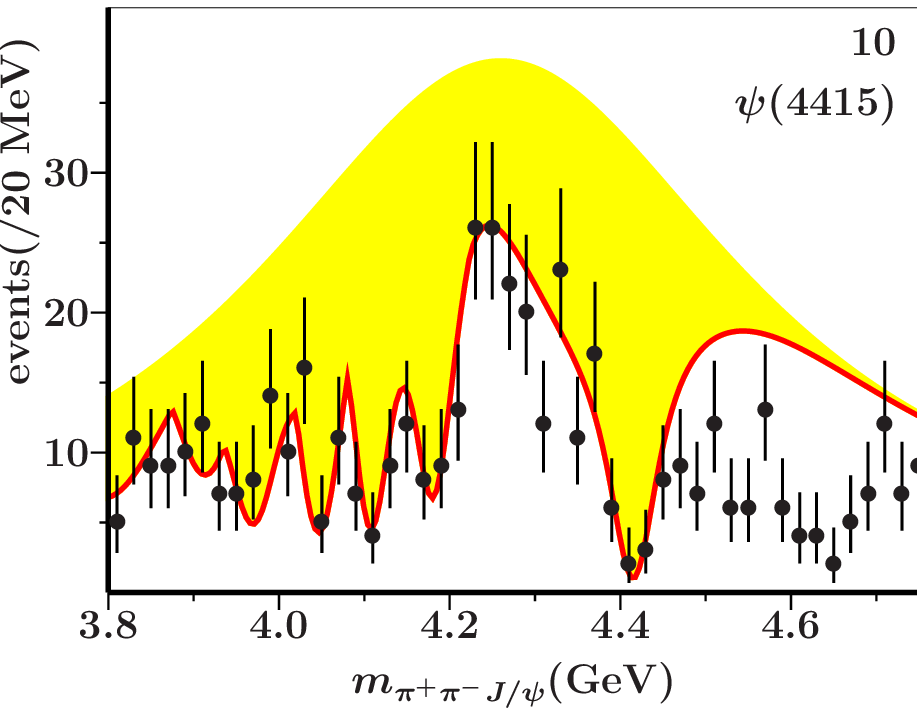} &
\includegraphics[height=116.48pt]{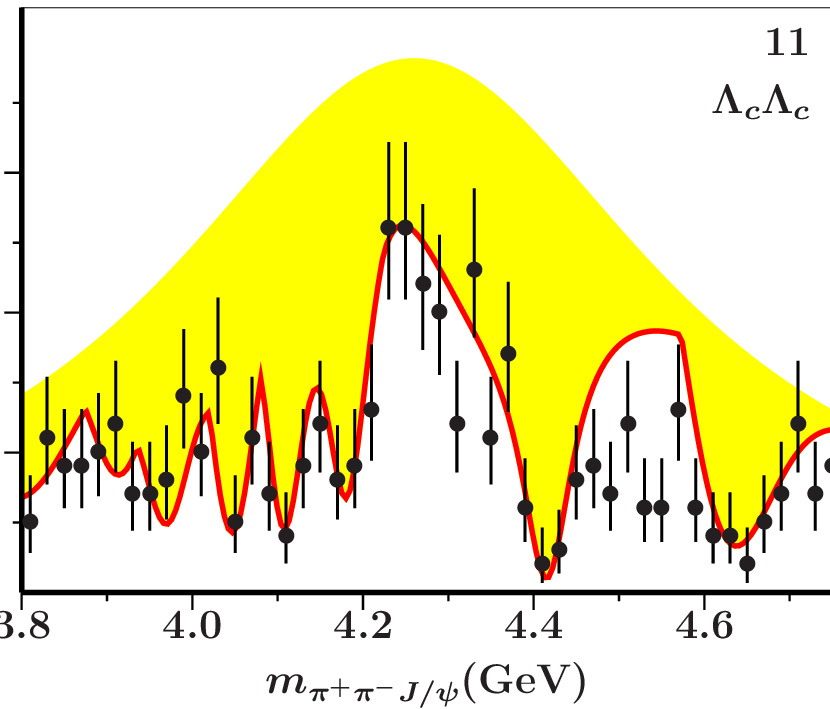} &
\includegraphics[height=116.48pt]{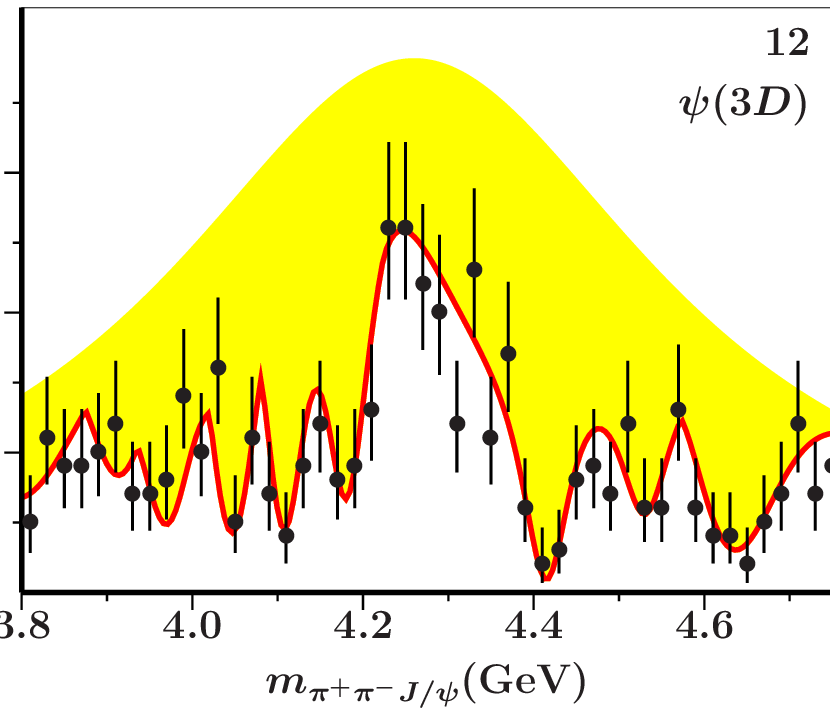} \\ [-20pt]
\end{tabular}
\end{center}
\caption{\small
A stepwise study of how the presumed $X(4260)$ signal
in $e^{+}e^{-}\to\pi^{+}\pi^{-}J/\psi$ \cite{PRL95p142001}
is depleted by OZI-allowed processes.
From upper left to lower right:
the presumed $X(4260)$ signal (1),
depletion by respective addition of
$D\bar{D}$ (2), $D\bar{D}^{\ast}$ (3), $D_{s}\bar{D}_{s}$ (4),
$D^{\ast}\bar{D}^{\ast}$ (5), $\psi (4040)$ (6), $D_{s}\bar{D}_{s}^{\ast}$ (7),
$\psi (4160)$ (8), $D_{s}^{\ast}\bar{D}_{s}^{\ast}$ (9), $\psi (4415)$ (10),
$\Lambda_{c}\bar{\Lambda}_{c}$ (11), and $\psi (3D)$ (12).
}
\label{x4260}
\end{figure*}
\addtocounter{figure}{-3}
\refstepcounter{figure}
Shortly after BaBar published its findings,
Zhu \cite{PLB625p212} proposed
a hybrid charmonium description of the phenomenon.
This proposal was later supported by
Close and Page \cite{PLB628p215},
whereas Kou and Pene \cite{PLB631p164}
advocated that the $X(4260)$ may be
a charmonium hybrid state with a magnetic constituent gluon.
Llanes-Estrada \cite{PRD72p031503}
suggested the $X(4260)$ to replace the $\psi(4415)$
as the (largely) $4S$ vector charmonium state.
He furthermore showed that the strong suppression of any $KKJ/\psi$
mode can be understood to be a consequence of chiral symmetry.
Curiously, the $K^{+}K^{-}J/\psi$ mode, with a significant branching
fraction, was in the meantime reported
by CLEO \cite{PRL96p162003} and Belle \cite{PRD77p011105}.
Maiani {\em et al.}
\cite{PRD72p031502}
proposed the $X(4260)$ to be the first orbital excitation
of a diquark-antidiquark state
($\left[ cs\right]\left[\bar{c}\bar{s}\right]$),
and in collaboration with Bigi \cite{PRD72p114016}
reminded us that the existence of four-quark configurations
might resolve the long-standing puzzle of
higher $\psi$ production in $B$ decays than expected, at momenta
below 1 GeV.
The tetraquark picture was also supported by the model calculation of
Ebert, Faustovo, and Galkin \cite{PLB634p214},
who obtained a value of 4244 MeV for
a bound state of a heavy-light diquark and antidiquark.
However, they excluded
a possible charm-strange diquark-antidiquark hypothesis
for the $X(4260)$, since its mass is predicted 200 MeV too heavy.
Liu, Zeng, and Li \cite{PRD72p054023}
suggested
a $\rho^{0}\chi_{c1}$ $S$-wave molecular picture for the $X(4260)$.
Alternatively, Yuan, Wang, and Mo \cite{PLB634p399}
proposed an $\omega\chi_{c1}$ $S$-wave molecular state.
A baryonium solution,
i.e., a bound state of a $\Lambda_{c}\bar{\Lambda}_c$ pair,
was proposed by Qiao \cite{PLB639p263}.
Recently,
deeply bound $S$-wave quasi-molecular charmed meson pairs,
bound by hundreds of MeVs, were suggested by
Close, Downum, and Thomas
\cite{PRD81p074033}
to describe the $X(4260)$ enhancement.
Hence, a plethora of --- often mutually contradicting --- explanations exist.
However, what puzzles us most is that all these approaches
completely ignore the phenomenon of prior interest to be addressed
concerning the $X(4260)$ enhancement, namely
the observation that the signal in $e^{+}e^{-}\to\pi^{+}\pi^{-}J/\psi$
is depleted exactly at the mass of the $\psi (4S)$
(see Fig.~\ref{B4260a}).

In Refs.~\cite{HEPPH0605317,ARXIV08111755,ARXIV09044351},
it was assumed that,
while the reaction $e^{+}e^{-}\to\pi^{+}\pi^{-}J/\psi$
is dominated by a peripheral, OZI-forbidden process,
in which a $\sigma$-like structure, i.e., $f_{0}(600)$ and/or
$f_{0}(980)$, is radiated off by the gluon cloud,
\begin{figure}[hbtp]
\begin{center}
\begin{tabular}{c}
\includegraphics[width=150pt]{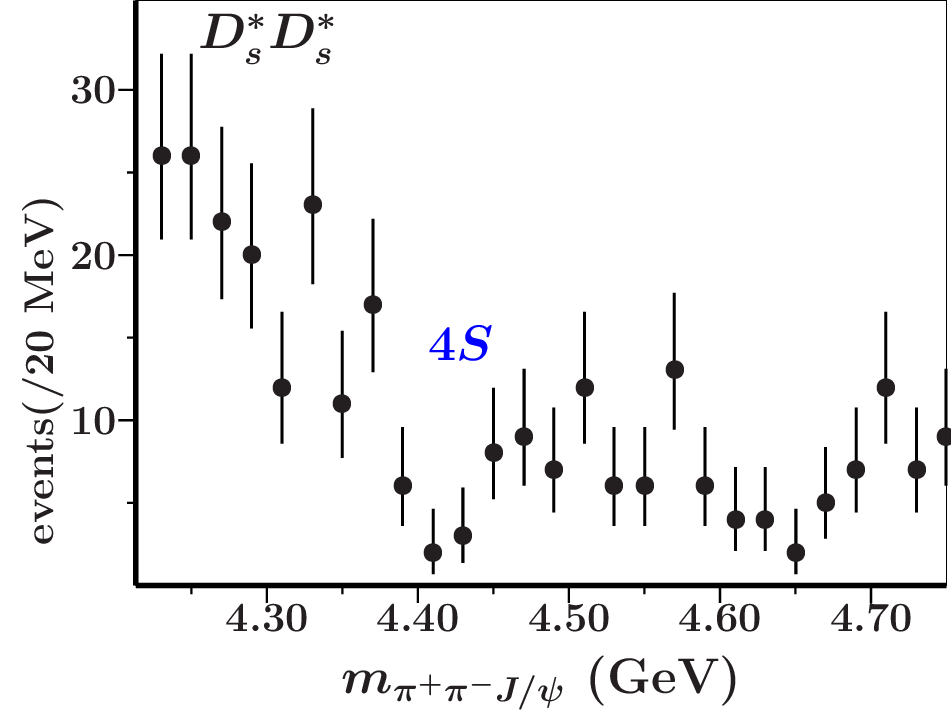}\\ [-20pt]
\end{tabular}
\end{center}
\caption{\small
Event distribution for
$e^{+}e^{-}\to\pi^{+}\pi^{-}J/\psi$,
as published by the BaBar Collaboration
\cite{PRL95p142001}.
}
\label{B4260a}
\end{figure}
\refstepcounter{figure}
the reaction $e^{+}e^{-}\to D^{\ast}\bar{D}^{\ast}$
is dominated by OZI-allowed quark-pair creation in the inner core
of the $c\bar{c}$ propagator.
Near a $c\bar{c}$ resonance,
the latter --- faster --- process dominates,
hence depleting the $\pi^{+}\pi^{-}J/\psi$ signal.
Actually, we may observe the lack of signal just above
all open-charm thresholds, that is,
$D\bar{D}$, $D\bar{D}^{\ast}$, $D^{\ast}\bar{D}^{\ast}$, $D_{s}\bar{D}_{s}$,
$D_{s}\bar{D}_{s}^{\ast}$, $D_{s}^{\ast}\bar{D}_{s}^{\ast}$,
$\Lambda_{c}\Lambda_{c}$,
and also at the known vector charmonium resonances
in the relevant invariant-mass region, viz.\
$\psi (4040)$, $\psi (4160)$, $\psi (4415)$, apart from the new $\psi (3D)$.
In Fig.~\ref{x4260} we depict the situation in a stepwise fashion.

We start from the Ansatz that the $X(4260)$ enhancement is given
by a broad structure peaking near 4.26 GeV.
Upon reconstructing the observed signal \cite{PRL95p142001},
we adjust the shape parameters of our Ansatz.
In Fig.~\ref{x4260}.1 we show the final shape,
which peaks at exactly 4.26 GeV and has a width of 700 MeV,
while Fig.~\ref{x4260}.2 depicts the fraction that we assume
to be consumed by the $D\bar{D}$ threshold enhancement.
In Fig.~\ref{x4260}.3 we add to this the fraction for $D\bar{D}^{\ast}$.
Each additional serving is indicated in the subsequent figures.
At the end of our exercise, we recover in Fig.~\ref{x4260}.12
what has been left for the process $e^{+}e^{-}\to\pi^{+}\pi^{-}J/\psi$,
which can be measured in experiment \cite{PRL95p142001}.

\begin{figure}[htbp]
\begin{center}
\begin{tabular}{c}
\includegraphics[height=150pt]{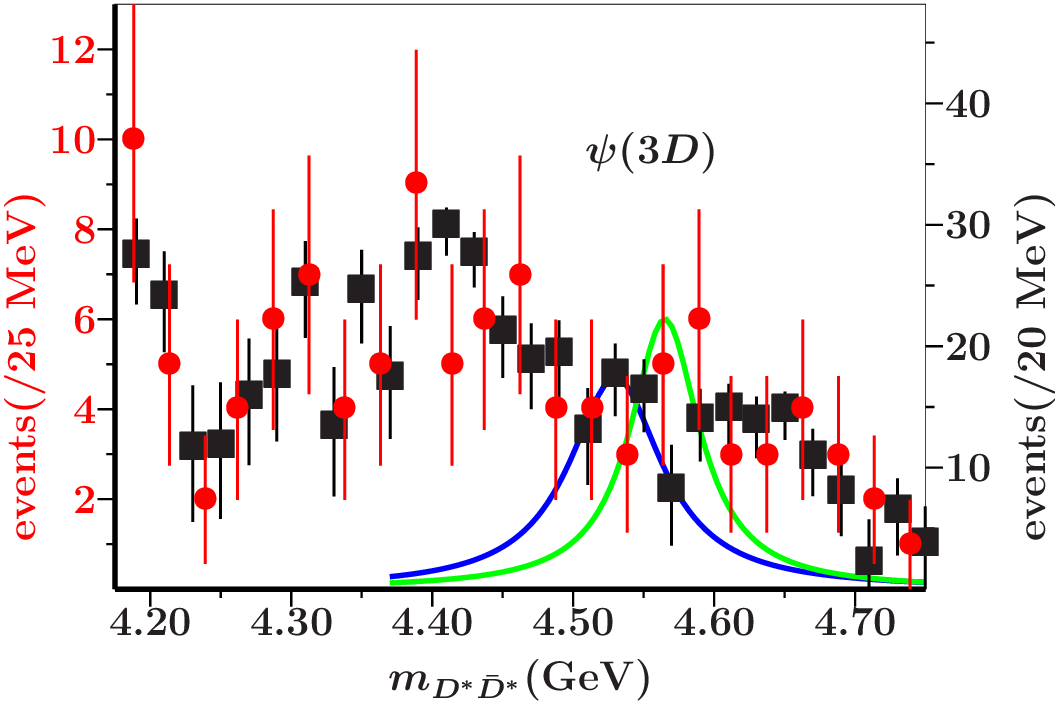}\\ [-20pt]
\end{tabular}
\end{center}
\caption[]{\small
BaBar data for $e^{+}e^{-}\to D^{\ast}\bar{D}^{\ast}$
\cite{PRD79p092001} ({\color{red}$\bullet$}),
and the missing signal in $e^{+}e^{-}\to\pi^{+}\pi^{-}J/\psi$,
\cite{PRL95p142001} ($\blacksquare$),
due to OZI-allowed decay processes
as shown in Fig.~\ref{x4260}.12
(see also Ref.~\cite{ARXIV09044351}).
The annotations at the vertical axis on the lefthand side
refer to the data of Ref.~\cite{PRD79p092001},
while those on the righthand side
concern the data of Ref.~\cite{PRL95p142001}.
The missing signal is adjusted in magnitude so as to be compared with
the $e^{+}e^{-}\to D^{\ast}\bar{D}^{\ast}$ data.
For each set of data,
$e^{+}e^{-}\to D^{\ast}\bar{D}^{\ast}$
\lijntje{0}{1}{0}
and the missing signal
\lijntje{0}{0}{1},
the determination of the resonance parameters
of the $\psi (3D)$ is shown.
}
\label{difference}
\end{figure}
Now, in order to judge whether our presumed shape of the $X(4260)$ enhancement
makes any sense, we shall compare it to production data for open-charm pairs.
To that end, in Fig.~\ref{difference}
we depict, in one and the same figure, BaBar production data
\cite{PRD79p092001}
for the open-charm reaction
$e^{+}e^{-}\to D^{\ast}\bar{D}^{\ast}$ ({\color{red}$\bullet$}),
as well as the differences between
the presumed shape of the $X(4260)$ enhancement
and the experimental data, also by BaBar\cite{PRL95p142001},
for $e^{+}e^{-}\to\pi^{+}\pi^{-}J/\psi$.
We have indicated in Fig.~\ref{difference}
how the magnitudes of the two signals are adjusted
in order to be comparable.
As a matter of fact, close to the $D^{\ast}\bar{D}^{\ast}$ threshold
(at 4.02~GeV) we cannot really compare the two data sets,
because the phase space factors
of $\pi^{+}\pi^{-}J/\psi$ and $D^{\ast}\bar{D}^{\ast}$
are very different at that energy.
However, from roughly 4.2~GeV upwards we may to some extent
ignore phase-space effects.

One observes in Fig.~\ref{difference}
that indeed the OZI-allowed signal of
$e^{+}e^{-}\to D^{\ast}\bar{D}^{\ast}$
is in very good agreement with the signal stemming
from the missing signal in $e^{+}e^{-}\to\pi^{+}\pi^{-}J/\psi$,
both sharing in detail their
maxima and minima as a function of invariant mass.
Consequently, in $e^{+}e^{-}\to\pi^{+}\pi^{-}J/\psi$
we appear to probe the very structure of the interior
of the $c\bar{c}$ propagator.
This is clearly demonstrated by our method
of accounting for all OZI-allowed decays,
depicted in Fig.~\ref{x4260}.1--12,
and the comparison we make in Fig.~\ref{difference}
between the direct measurement of the $c\bar{c}$ structure
in $e^{+}e^{-}\to D^{\ast}\bar{D}^{\ast}$,
and the indirect measurement extracted from
the OZI-forbidded process $e^{+}e^{-}\to\pi^{+}\pi^{-}J/\psi$.

In Fig.~\ref{difference} we indicate
two independent methods for determining the $\psi (3D)$
resonance parameters.
First,
we observe the contribution of the $\psi (3D)$ resonance
in arriving at Fig.~\ref{x4260}.12,
starting from Fig.~\ref{x4260}.11,
where the contribution for
the $\Lambda_{c}\bar{\Lambda}_{c}$ threshold enhancement
is depicted.
Its resonance parameters are given by
(central mass, width) = (4.53 GeV, 80 MeV).
Second,
for the $e^{+}e^{-}\to D^{\ast}\bar{D}^{\ast}$ data, we find
(4.565 GeV, 60 MeV).
Differences are not unexpected,
as each channel reflects the resonance pole
through a different shape.
Moreover, the experimental data leave enough room for some uncertaincy.
We baptize this resonance as $\psi (3D)$, since it comes out exactly
in the mass interval predicted a long time ago
\cite{PRD21p772,PRD32p189}
for the $c\bar{c}$ $\psi (3D)$ state.
In the following, we shall present further evidence
for its existence in the mass range 4.53--4.58 GeV.
Note, however, that more recent predictions,
aimed at accomodating $XYZ$ states in the $c\bar{c}$ spectrum,
obtain $\psi (3D)$ masses that are some 100~MeV lower,
viz.\ 4477 MeV \cite{PAN72p638},
4455 MeV \cite{PRD72p054026},
and 4426 MeV \cite{PRD78p114033}.

\begin{figure}[htbp]
\begin{center}
\begin{tabular}{ccc}
\hspace{-3pt}\includegraphics[height=100pt]{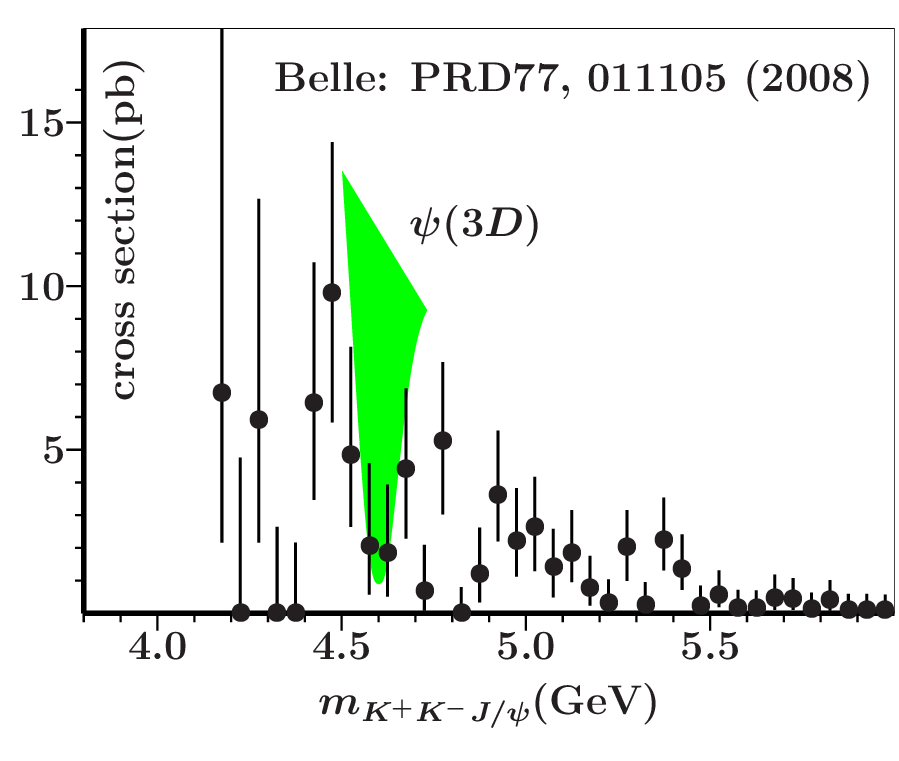} &
&
\includegraphics[height=100pt]{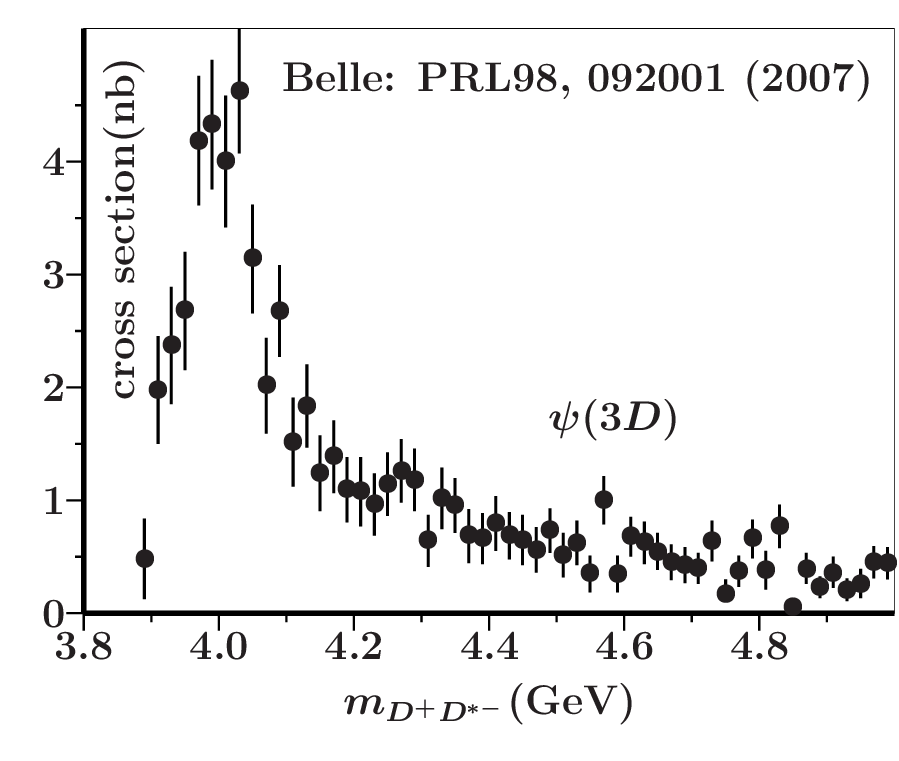}\\ [-20pt]
\mbox{}\hspace{100pt}{\bf a} & &
\mbox{}\hspace{106pt}{\bf b}\\ [3pt]
\hspace{-3pt}\includegraphics[height=100pt]{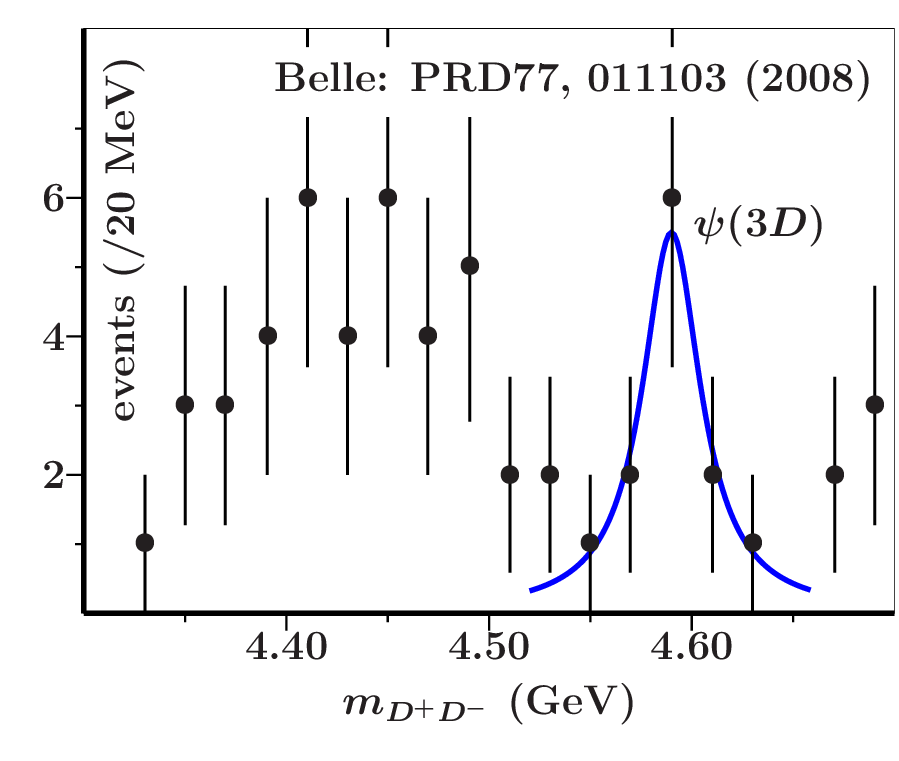} &
&
\includegraphics[height=100pt]{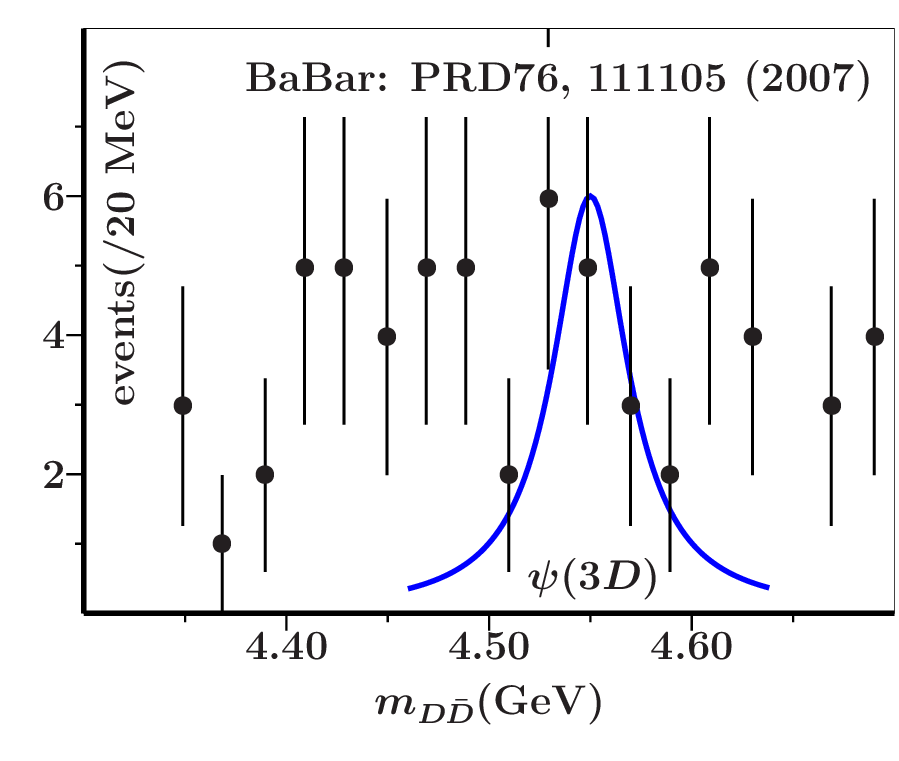}\\ [-20pt]
\mbox{}\hspace{100pt}{\bf c} & &
\mbox{}\hspace{106pt}{\bf d}\\ [-10pt]
\end{tabular}
\end{center}
\caption[]{\small
The missing data ({\color{green} shaded area})
in the $e^{+}e^{-}\to K^{+}K^{-}J/\psi$ signal
(a: Belle \cite{PRD77p011105})
due to the $\psi (3D)$ resonance.
Three hints for the $\psi (3D)$ resonance
in
$e^{+}e^{-}\to D^{+}\bar{D}^{\ast -}$
(b: Belle \cite{PRL98p092001}),
$e^{+}e^{-}\to D^{+}\bar{D}^{-}$
(c: Belle \cite{PRD77p011103}),
and $e^{+}e^{-}\to D\bar{D}$
(d: BaBar \cite{PRD76p111105}), respectively.
}
\label{psi3Ds}
\end{figure}

We observe a modest peak at 4.57 GeV (see Fig~\ref{psi3Ds}b)
in the  Belle \cite{PRL98p092001} $e^{+}e^{-}\to D^{+}\bar{D}^{\ast -}$
cross section. However, comprising a mere three data points,
its width can only be very roughly estimated to be of the order of 50 MeV.
Then, we observe that Belle data
for $e^{+}e^{-}\to D^{+}\bar{D}^{-}$ \cite{PRD77p011103}
and BaBar data for $e^{+}e^{-}\to D\bar{D}$ \cite{PRD76p111105}
qualitatively agree with one another
(see Fig~\ref{psi3Ds}c (Belle) and Fig~\ref{psi3Ds}d (BaBar)),
namely in displaying a relatively broad bump, the $\psi(4S)$,
in the mass interval 4.4--4.5 GeV,
another peak in the mass interval 4.5--4.6 GeV,
which is more conspicuous in the Belle data,
and the onset of the enhancement due to the opening
of the $\Lambda_{c}^{+}\Lambda_{c}^{-}$ channel at 4.573 GeV.
As far as these data allow such a treatment,
we deduce (4.59 GeV, 35 MeV) for the resonance parameters
of the $\psi (3D)$ from the Belle data,
and (4.55 GeV, 45 MeV) from the BaBar data.
Finally, we observe an almost complete depletion of
the $e^{+}e^{-}\to K^{+}K^{-}J/\psi$ signal
in the  Belle \cite{PRD77p011105} cross section
at the position of the $\psi (3D)$ resonance
(see Fig~\ref{psi3Ds}a),
preceded by a total depletion
at the position of the $\psi (4S)$ resonance,
and followed by a similar effect due to the opening
of the $\Lambda_{c}^{+}\Lambda_{c}^{-}$ channel.

In view of the above, we may quite safely conclude
that the $\psi (3D)$ charmonium state has been observed.
However, the data do not allow a rigorous determination
of its resonance parameters, and only indicate a range of
4.53--4.58 GeV for the central mass and 40--70 MeV for the width.
Thus, the open-charm decay width of the $\psi (3D)$ seems
somewhat smaller than naively expected.
However, a thorough discussion of this issue lies outside the scope
of the present paper.
Finally, we have shown that the $X(4260)$ enhancement,
when carefully analysed,
contains a wealth of information on
the properties of the $c\bar{c}$ propagator.
All known charmonium vector enhancements,
resonances, and threshold openings
have been identified by us in the broad structure
that reveals itself in the OZI-forbidden process
$e^{+}e^{-}\to\pi^{+}\pi^{-}J/\psi$.
The broad $X(4260)$ enhancement itself, which
does not seem to classify as a vector $c\bar{c}$ resonance,
reminds of the two-pion shape of the $f_{0}(600)$, also because of
its apparently preferential production mechanism, involving two pions
with plenty of phase space.

We are grateful for the precise measurements
and data analyses of the BaBar and Belle Collaborations,
which made the present analysis possible.
This work was supported in part by the {\it Funda\c{c}\~{a}o para a
Ci\^{e}ncia e a Tecnologia} \/of the {\it Minist\'{e}rio da Ci\^{e}ncia,
Tecnologia e Ensino Superior} \/of Portugal, under contract
CERN/FP/ 109307/2009,
and by the {\it Ministerio de Ciencia y Tecnolog\'{\i}a de Espa\~{n}a}
under contract FPA2007/\-65748.

\newcommand{\pubprt}[4]{#1 {\bf #2}, #3 (#4)}
\newcommand{\ertbid}[4]{[Erratum-ibid.~#1 {\bf #2}, #3 (#4)]}
\def\AP{Ann.\ Phys.}
\def\PAN{Phys.\ Atom.\ Nucl.}
\def\PLB{Phys.\ Lett.\ B}
\def\PRD{Phys.\ Rev.\ D}
\def\PRL{Phys.\ Rev.\ Lett.}


\begin{thebibliography}{31}
\bibitem{PRD79p092001}
B.~Aubert  [The BABAR Collaboration],
\pubprt{\PRD}{79}{092001}{2009}.

\bibitem{PLB667p1}
C.~Amsler {\it et al.} \/[Particle Data Group Collaboration],
\pubprt{\PLB}{667}{1}{2008}.

\bibitem{AP323p1215}
E.~van Beveren and G.~Rupp,
\pubprt{\AP}{323}{1215}{2008}.

\bibitem{PRD21p772}
E.~van Beveren, C.~Dullemond, and G.~Rupp,
\pubprt{\PRD}{21}{772}{1980}
\ertbid{\ D}{22}{787}{1980}.

\bibitem{PRD32p189}
S.~Godfrey and N.~Isgur,
\pubprt{\PRD}{32}{189}{1985}.

\bibitem{PRL95p142001}
B.~Aubert {\it et al.}  [BABAR Collaboration],
\pubprt{\PRL}{95}{142001}{2005}.

\bibitem{PRL96p162003}
T.~E.~Coan {\it et al.}  [CLEO Collaboration],
\pubprt{\PRL}{96}{162003}{2006}.

\bibitem{PRL99p182004}
C.~Z.~Yuan {\it et al.}  [Belle Collaboration],
\pubprt{\PRL}{99}{182004}{2007}.

\bibitem{PRL98p212001}
B.~Aubert {\it et al.}  [BABAR Collaboration],
\pubprt{\PRL}{98}{212001}{2007}.

\bibitem{PRL99p142002}
X.~L.~Wang {\it et al.}  [Belle Collaboration],
\pubprt{\PRL}{99}{142002}{2007}.

\bibitem{PLB625p212}
S.~L.~Zhu,
\pubprt{\PLB}{625}{212}{2005}.

\bibitem{PLB628p215}
F.~E.~Close and P.~R.~Page,
\pubprt{\PLB}{628}{215}{2005}.

\bibitem{PLB631p164}
E.~Kou and O.~Pene,
\pubprt{\PLB}{631}{164}{2005}.

\bibitem{PRD72p031503}
F.~J.~Llanes-Estrada,
\pubprt{\PRD}{72}{031503}{2005}.

\bibitem{PRD77p011105}
C.~Z.~Yuan {\it et al.}  [Belle Collaboration],
\pubprt{\PRD}{77}{011105}{2008}.

\bibitem{PRD72p031502}
L.~Maiani, V.~Riquer, F.~Piccinini and A.~D.~Polosa,
\pubprt{\PRD}{72}{031502}{2005}.

\bibitem{PRD72p114016}
I.~Bigi, L.~Maiani, F.~Piccinini, A.~D.~Polosa and V.~Riquer,
\pubprt{\PRD}{72}{114016}{2005}.

\bibitem{PLB634p214}
D.~Ebert, R.~N.~Faustov and V.~O.~Galkin,
\pubprt{\PLB}{634}{214}{2006}.

\bibitem{PRD72p054023}
X.~Liu, X.~Q.~Zeng and X.~Q.~Li,
\pubprt{\PRD}{72}{054023}{2005}.

\bibitem{PLB634p399}
C.~Z.~Yuan, P.~Wang and X.~H.~Mo,
\pubprt{\PLB}{634}{399}{2006}.

\bibitem{PLB639p263}
C.~F.~Qiao,
\pubprt{\PLB}{639}{263}{2006}.

\bibitem{PRD81p074033}
F.~Close, C.~Downum and C.~E.~Thomas,
\pubprt{\PRD}{81}{074033}{2010}.

\bibitem{HEPPH0605317}
E.~van Beveren and G.~Rupp,
arXiv:hep-ph/0605317.

\bibitem{ARXIV08111755}
E.~van Beveren and G.~Rupp,
in Proceedings {\it Bled Workshops in Physics},
Vol.~9, no.~1, pp 26-29 (2008).

\bibitem{ARXIV09044351}
E.~van Beveren and G.~Rupp,
arXiv:0904.4351 [hep-ph].

\bibitem{PAN72p638}
A.~M.~Badalian, B.~L.~G.~Bakker, and I.~V.~Danilkin,
\pubprt{\PAN}{72}{638}{2009}.

\bibitem{PRD72p054026}
T.~Barnes, S.~Godfrey and E.~S.~Swanson,
\pubprt{\PRD}{72}{054026}{2005}.

\bibitem{PRD78p114033}
J.~Segovia, A.~M.~Yasser, D.~R.~Entem and F.~Fernandez,
\pubprt{\PRD}{78}{114033}{2008}.

\bibitem{PRL98p092001}
K.~Abe {\it et al.}  [Belle Collaboration],
\pubprt{\PRL}{98}{092001}{2007}.

\bibitem{PRD77p011103}
G.~Pakhlova {\it et al.}  [Belle Collaboration],
\pubprt{\PRD}{77}{011103}{2008}.

\bibitem{PRD76p111105}
B.~Aubert  [BaBar Collaboration],
\pubprt{\PRD}{76}{111105}{2007}.
\end{thebibliography}
\end{document}